# OpenDigger：面向开源协作数字生态的数据挖掘与信息服务系统

X-lab 开放实验室，2023 年 7 月

**摘要** 开源软件的大规模发展与普及的同时也构筑了一个开源开发与协同的生态系统，在这个系统中，个人与组织协同开发所有人都可以使用的高质量软件。以 GitHub 为代表的社会化协作平台进一步促进了大规模、分布式、细粒度的代码协作与技术社交，无数开发者每天在其上提交代码、评审代码，报告 bug、或提出新的功能请求，使得全开放的协作开发过程所产生的海量行为数据具备极大价值。本文设计并实现了一个面向开源协作数字生态的一站式数据挖掘与信息服务系统 OpenDigger，目标是构建开源领域的数据基础设施，促进开源生态的持续发展。OpenDigger 系统的指标体系和分析模型可以从宏观到微观不同层面，挖掘开源数字生态中的各种知识，并通过统一的信息服务接口为政府、企业、基金会、个人等不同用户群体提供各类开源信息服务。作为一个开源生态场景下的新型信息服务系统，本文通过落地于真实场景中的若干实例（包括产品、工具、应用、课程等），说明了 OpenDigger 中的指标体系与模型的有效性，在算法侧和业务侧都具备重要且多元的应用实践。

**OpenDigger: Data Mining and Information Service System for Open Collaboration Digital Ecosystem**

**Abstract** The widespread development and adoption of open-source software have built an ecosystem for open development and collaboration. In this ecosystem, individuals and organizations collaborate to create high-quality software that can be used by everyone. Social collaboration platforms like GitHub have further facilitated large-scale, distributed, and fine-grained code collaboration and technical interactions. Countless developers contribute code, review code, report bugs, and propose new features on these platforms every day, generating a massive amount of valuable behavioral data from the open collaboration process. This paper presents the design and implementation of OpenDigger, a comprehensive data mining and information service system for open collaboration in the digital ecosystem. The goal is to build a data infrastructure for the open-source domain and promote the continuous development of the open-source ecosystem. The metrics and analysis models in the OpenDigger system can mine various knowledge from the macro to micro levels in the open-source digital ecosystem. Through a unified information service interface, OpenDigger provides various open-source information services to different user groups, including governments, enterprises, foundations, and individuals. As a novel information service system in the open-source ecosystem, this paper demonstrates the effectiveness of the metrics and models in OpenDigger through several real-world scenarios, including products, tools, applications, and courses. It showcases the significant and diverse practical applications of the metrics and models in both algorithmic and business aspects.

近年来，开源软件的持续发展得到了全球社会的极大关注，在全球数字化创新与转型中的地位、在不同规模组织数字主权中的地位已经得到了广泛共识[1]。开源软件不仅构成了现代数字文明的基石，开源协作对人类数字文明的发展也起到了巨大推动作用。基于 Git 的分布式协作成为全球范围内最主要的开源创新模式，无数个开源社区在其上孕育而生，其背后海量的开发者行为数据蕴含了大量的个体贡献规律、群体协作模式、社区健康状况、生态发展趋势、以及商业战略价值。随着 Git 开发模式的标准化，以 GitHub 为代表的平台把围绕一个开源项目的关键环节打包在了一起，成为数字产品创作与生成的流水线，而这条流水线上的所有过程数据，都因开源的属性而变得透明，进而可分析、可理解、可利用。对这些数据的采集与存储、指标定义与建模、数据挖掘与分析可以帮助研究者和从业者了解个体贡献与群体协同模式，并进一步理解开源生态的发展特征[1]。

与此同时，如何使开源软件生态持续健康发展这个课题得到了来自学术界与工业界的广泛研究。通过数据与量化的手段来研究开源软件生态的持续发展开始成为全球研究学者的一项重要活动，包括项目/仓库层面、社区/组织层面、以及生态/社会层面。像 GitHub 这样全球化的开源生态网络，蕴含了非常丰富的数据挖掘问题。例如，由 Git 协作数据构建的网络，可以看作是一个典型的社会技术系统[2-3]（Socio-technical System），其中的开发者、代码、仓库、commit、制品、组织、issue、PR、社区、供应链、生态圈等，形成了一个极其复杂、且不断时序变化的复杂信息网络[4]。

开源生态数据存在结构复杂性（代码数据、时序数据、关联数据、NLP 数据等）与语义丰富性（代码语义、社交语义、协作语义、技术语义等）的特点。近期各种安全事件的层出不穷，又让全球软件供应链问题浮出水面。开源代码本身的价值毋庸置疑，类似 World of Code (WoC) [5]这样的"大代码"工程系统也是非常基础性的。但开发者作为开源生态的核心，在开源协作和开源项目生产过程中所产生的行为数据和过程性数据也至关重要，所谓“社区大于代码（Community over code）”[6]。

基于此，本文设计并实现了以开源协作数据为核心的开源软件数字生态信息服务系统 OpenDigger，图 1 展示了 OpenDigger 长期所面向的开源数字生态、OpenDigger 目前的系统模块和 OpenDigger 已落地与潜在的下游应用场景。OpenDigger 信息系统可以满足:（1）产业界对开源生态洞察的需求。它提供了一种全面、系统化的方法来收集、清洗、分析开发者行为数据，帮助从业者了解开发者的行为模式和协作关系，以及开源项目的健康度与价值潜力；（2）学术界对开源生态研究的需求。它拥有的海量的全域开源数据是开展具有共识性的领域问题、数据问题和数学问题的数据基础设施和底座。

本文的主要贡献包括以下部分：

1）设计并实现了开源生态数据挖掘与信息服务系统 OpenDigger。它包括软件开发协作数据、度量指标、度量模型、实现算法、标签数据、分析工具、以及社区分析和下游应用案例集。

2）在 OpenDigger 的数据挖掘模块，我们提出并实现了两个分析模型：基于统计模型的开源开发者活跃度和基于异质信息网络降维的开源项目影响力，基于这两个模型的数据挖掘能够提供启发于业务决策的洞察分析。

3）落地了 OpenDigger 在多个业务场景下的真实应用，证明了在开源数字生态业务侧的数据消费能力，以及 OpenDigger 作为开源生态数据信息系统的必要性和价值。

## 1 相关工作

### 1.1 开发者协作行为数据挖掘

开发者协作行为数据分析是软件工程的研究热点之一。袁等人[7]通过构造提交者网络发现了项目开发过程中主要开发人员频繁参与协同行为等现象.随后，李等人[8]分析了 Github 上开发者贡献度的影响因素,进一步挖掘了开发者所属地域与开发者协作之间的关系。不少研究者通过研究 GitHub 的社交特征来评估项目并挖掘开发者之间潜在的合作机会[9]。Marlow 等人[10]观察到开发人员一般使用开发者个人简介中的信息（例如，技能和人物关系）以形成对开发者或者项目的初步了解，其中对开源社区中如何通过拉取请求来改进成员贡献的认可度尤其关注。Tsay 等人[11]通过对开发者协作行为研究后发现，开发者的个人技能和其社交关系都会影响个人贡献的评估。McDonald 和 Goggins 指出[12]，GitHub 平台提供的协作功能是开源项目中的开发人员数量和贡献不断增加的主要原因。

### 1.2 开源软件数据挖掘

软件数据挖掘是国际软件工程领域最为活跃的

研究方向之一[13]。早期，Samoladas 等人[14]提出一种层次度量模型（SQO-OSS）,该模型通过挖掘软件开发过程与社区运营的数据，基于度量值的自动计算和一组预定义的度量配置文件的相关性来评估源代码和社区文档，但评估指标相对复杂，较难实现自动化. 基于此，Bauer[15]等人提出了一种基于质量特征的增量分布式计算框架，该框架通过对软件代码静态分析，可以实时对软件质量进行评估. 然而这些方法仅关注软件的内部质量,无法对软件的可用性和性能等方面进行度量。Zou 等人[16]通过挖掘软件用户的评论文本数据,提出了一种基于互联网用户评论的软件质量度量方法,通过对文本的情感分析,对软件质量的不同方面进行综合评估. 当前，随着以开源中国、StackOverflow 等国内外开源社区的逐步发展，通过挖掘社区用户对软件各方面的问答文本可以获取更多针对开源软件发展有价值的信息。Allamanis 等人[17]通过分析 StackOverflow 中的问答文本发现,该社区中的问题涵盖了不同编程语言、开发平台且涉及到各种类型的问题. Hen β 等人[18]通过结合文本挖掘和自然语言处理技术，提出了一种开源社区中自动提取常见问答的方法，以问卷的形式验证了方法的实用性。此外，学者们还从软件研发的不同角度以数据挖掘的方式对其展开研究，其中包含代码注释自动生成[19]、API 关联文档获取[20]、知识分享社区与开发环境的集成[21]等研究方向.

### 1.3 开源生态信息系统

开源生态数据系统可以为开发者提供开源领域的数据服务，推动开源软件的可持续发展。GHArchive[1]是一个开源的数据服务项目，用于记录公共 GitHub 时间轴，对其进行存档，并使其易于访问以进行更进一步地分析。它可以获取所有的 GitHub events 信息并存储在一组 JSON 文件中，以便根据需要下载进行脱机处理。该数据集每小时自动更新一次，可以在几秒钟内对整个数据集运行任意类似 SQL 的查询。和 GHArchive 类似，GHTorrent[22]项目也可以用来监视 Github 公共事件时间表信息。对于每个事件，它都详尽地检索其内容和相互依赖性。然后将结果 JSON 的信息存储到 MongoDB 数据库，同时还将其结构提取到 MySQL 数据库中。以上研究更多以提供公共数据为主，未考虑开源软件生态系统中项目或开发者之间的互相引用。World of Code[5]是一个数据量大且频繁更新的版本控制数据集，该数据集包含了作者、项目、提交信息、Blob 等多个实体之间的交叉引用信息，提供了更正、扩充、查询和分析这些数据的能力。OSS Insight[2]通过分析数十亿的 GitHub 事件数据，借助 TiDB 提供了实时查看和分析 GitHub 趋势的能力。它还提供了对单个 GitHub 仓库和开发人员的深入分析的能力。以上研究未以协作行为数据为核心，形成具有多项评价指标的数据服务系统。本文以数据服务为驱动，结合开源生态现有活动数据、协作行为数据、度量指标等实现了开源生态数据信息系统与挖掘平台，并从开发者和项目不同的角度，基于统计模型和异质信息网络分别形成了统一的评价指标。

## 2 OpenDigger 总体架构

OpenDigger 从开源生态大数据中提取有价值的信息和模式，目前所支持的业务场景主要为开源社区治理与运营场景下的数据建模、分析与洞察；所支持的开源领域数据是指由开源业务场景中的各类开源活动所产生的数据，目前主要为全域的开发者协作行为数据（例如 Git-log、问答平台数据、代码变更请求数据）。一个完整的 OpenDigger 信息服务生态系统包括了图 1 所示的三个主要部分：

**开源数字生态**：由软件生态与人类社会组成，社会组织与广大开发者通过软件开发活动构建各种各样的软件产品与服务，而这些产品和服务进而又服务于各种人类社会的需求。随着全球数字化转型的加速，这些开发活动与服务需求呈爆炸性增长的趋势，而开源软件开发模式更是进一步促进了这种全球化大规模协作的发展；

**数据挖掘与信息服务系统**：对上述全球化软件开发活动的各类数据（内容数据、行为数据等）进行采集、分析、挖掘、服务的过程，即 OpenDigger 的主体部分，第 2、3 节详细论述；

**应用**：对 OpenDigger 产生的信息进行实际消费，指导并推动开源生态的持续发展。为包括开源生态洞察报告、开源项目可视化插件与看板、开源社区激励榜单、企业开源治理大屏等应用提供信息服务接口。本文在第 4 部分选取若干典型进行详细描述。

在技术架构上，OpenDigger 利用开放式的开源数据集，与 OLAP 数据库 Clickhouse 和图数据库 Neo4j 进行集成，用于进行高效的统计分析和关联分析。在内置的丰富指标集的基础上，OpenDigger 支持在 Jupyter Notebook 环境下进行在线交互式开发，并能够进行线上部署的定时数据生成任务。

[1] https://www.gharchive.org/

[2] https://ossinsight.io/

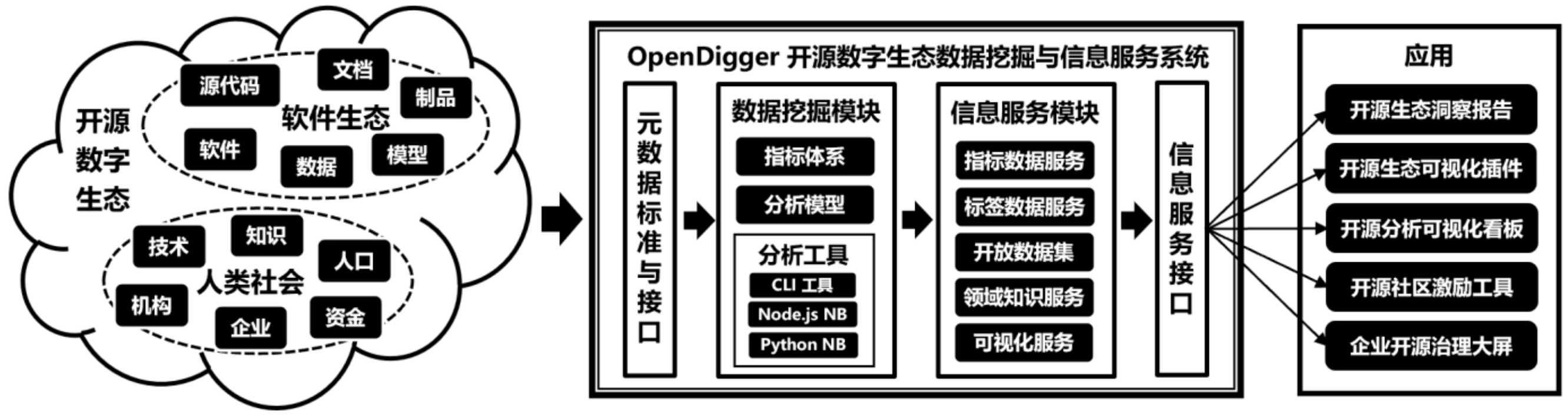

Fig. 1 OpenDigger Data Mining and Information Service Ecological Structure

图 1 OpenDigger 数据挖掘与信息服务生态结构

同时，OpenDigger 还开发了命令行界面（CLI）的数据查询分析工具。通过满足数据开发人员快速开发数据指标的需求，OpenDigger 还能够为下游应用提供丰富的实时数据支持。OpenDigger 的技术架构如图 2 所示。

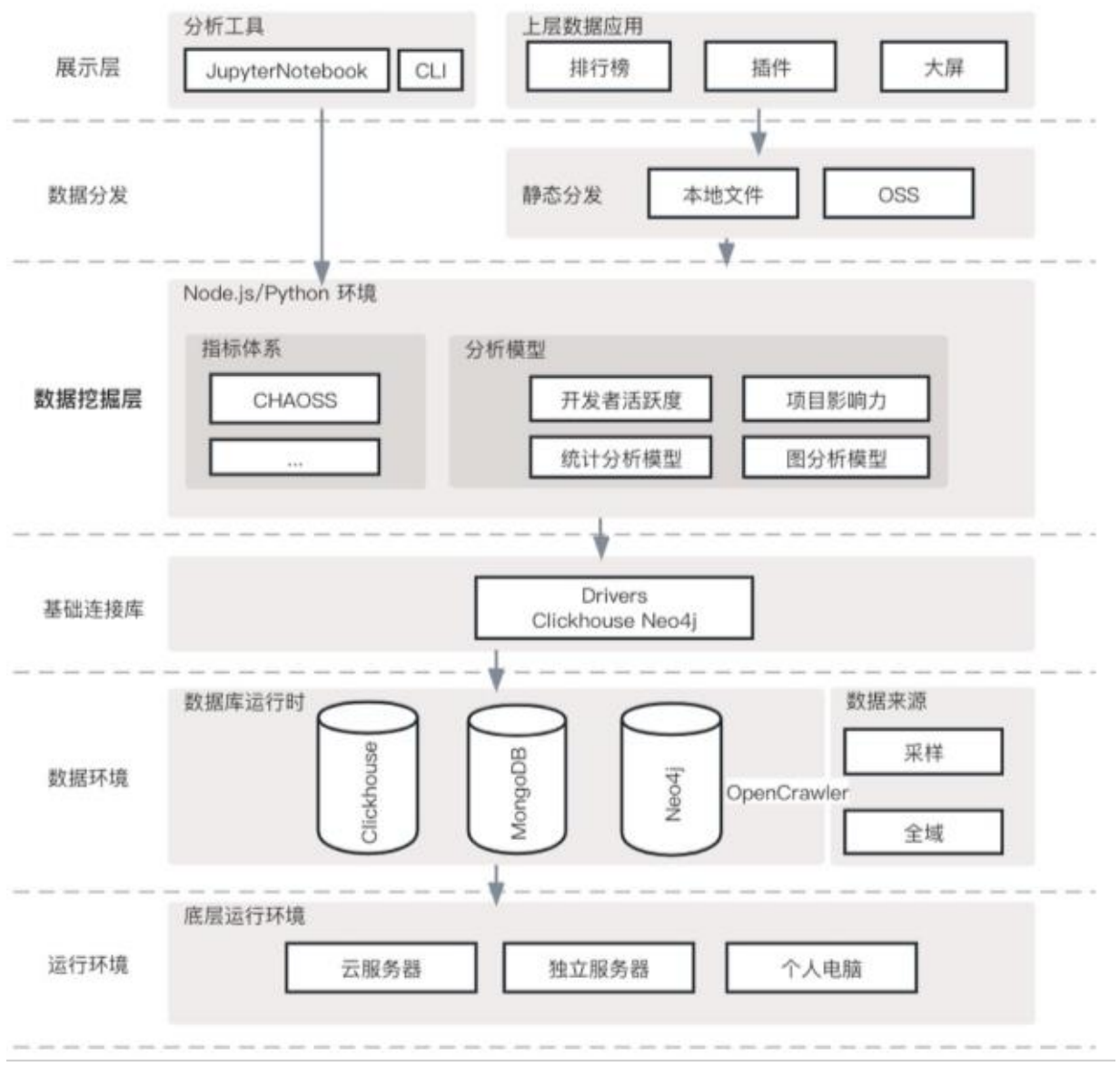

Fig. 2 OpenDigger Technology Architecture

图 2 OpenDigger 技术架构

## 2.1 数据采集与存储

OpenDigger 实现了 OpenCrawler 数据采集服务来配置各类开源数字生态数据的采集任务，并通过定时任务管理器进行调度来保证任务的自动化运行，实现数据的持续采集与存储。截至 2023 年 6 月，已采集到的各类开源生态数据的情况为：

- GitHub 协作行为日志数量 58 亿条
- NPM 制品库包数据 247 万条
- Go 语言模块数据 102 万条
- Maven 制品库包数据 55.9 万条
- PyPI 制品库包数据 44.9 万条
- NuGet 制品库包数据 36.4 万条
- Cargo 制品库包数据 12.1 万条
- CVE 安全漏洞数据 15.5 万条
- StackOverflow 问答帖 2413 万条

在本文中，将以开源协作数字生态为主体，重点介绍 OpenDigger 如何对 GitHub 协作行为日志数据进行采集，存储，建模，分析并提供数据服务与支持下游应用。

### 2.1.1 协作日志数据源

开源协作行为数据包含了开发者在代码活动以外依托平台进行问题交流、团队管理、版本发布、变更评审等软件开发管理过程中产生的行为数据。GitHub 作为全球最大的开源协作平台，拥有海量的开发者行为日志数据，是研究开源生态系统下开发者协作的优质数据源。

如图 3 展示了 GitHub 平台上的对象与对象之间的关系，开发者、仓库、与组织通过在文件级别（Files）、Issue 级别和 Pull Request 级别上的协作关联在一起，而每一次协作的行为都通过事件流的形式在时间轴上被记录下来。

GitHub 共定义了 17 种可被公开事件数据流采集的事件类型，不同的事件类型表征了开发者与项目之间的不同链接关系，即某开发者在某时间点向某仓库进行了某种操作，不同的事件类型会带有与其事件类型相对应的特有数据信息。在开发协作过程中较常出现的事件类型是 ForkEvent、WatchEvent、IssuesEvent、PREvent、IssueCommentEvent、PRReviewCommentEvent。对于这类开发者协作日志数据，本文选择 GHArchive 作为数据源。GHArchive 将事件流数据按小时进行汇总并归档，每个归档数据

文件都包含了一个小时内通过 GitHub 公开事件流 API 采集的 JOSN 格式数据，并最终打包一个 JSONL 的 gzip 压缩数据文件供下游使用。

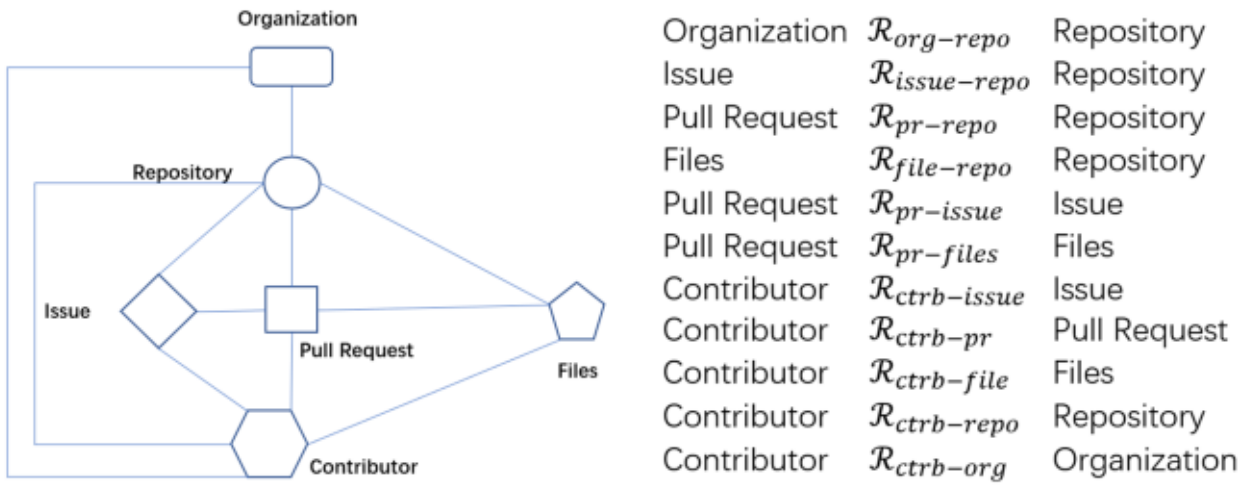

Fig. 3 GitHub Data Objects and Relations

图 3 GitHub 平台的数据对象与关系

### 2.1.2 数据采集服务

本文构建了一个名为 OpenCrawler 的数据采集服务框架，通过多个内置模块可以对数据采集过程进行定制与精细化控制，从而使应用层可以高效实现针对不同数据源的持续采集服务。OpenCrawler 使用由阿里巴巴开源的应用框架 Egg.js 构建，在基础框架之上提供了如下一些模块:

- **HTTP 请求管理器**：HTTP 请求管理器模块主要帮助业务层管理 HTTP（HTTPS）请求，可以支持请求洋葱模型、并发控制、自定义失败策略、动态添加请求、动态 IP 代理等能力。
- **Token 请求管理器**：Token 管理模块用于对各平台通用的访问频率受限的 token 进行池化管理，该模块会维护 token 池内每个 token 的当前状态，包括 token 是否可用、当前剩余的配额次数、下次配额的刷新时间点等，可用于 token 的精确管理。该模块提供了获取 token 的接口，可以从当前可用 token 中随机获取一个返回用于请求使用，可配合 HTTP 操作器模块的预处理接口进行 token 置入。
- **动态 IP 代理模块**：动态 IP 代理模块与 HTTP 操作器模块配合，用于在请求时设置动态 IP 代理。该模块支持多种主流的 IP 代理平台，可自动更新 IP 代理池，自动进行白名单配置，并在 HTTP 操作器模块中根据返回内容预判断是否为代理失效，若为代理失效，则会自动更新代理 IP 并重新发起请求。
- **大文件下载管理器**：虽然静态文件可以通过 HTTP 操作器进行下载，但当网络不稳定、下载文件体积较大时，可能频繁出现下载错误和重新下载，不仅耗时较长，而且内容先加载到内存再进行持久化可能导致内存占用过大。大文件下载管理器专门针对较大静态文件的并发下载需求，提供了直接面向磁盘的流式持久化能力，并提供了断点续传的能力，避免大文件下载过程中由于网络抖动导致的频繁重新下载。
- **多核任务管理器**：由于 OpenCrawler 使用 Egg.js 作为底层框架和 Node.js 的单线程语言模型，对于一些业务层需要进行高并发 CPU 密集型任务处理的需求，需要提供一套可以将业务分发到多核 CPU 上并行处理的机制，如内存中的流式解压缩、文件格式校验、数据格式化等 CPU 密集型任务。该模块支持自动识别当前硬件环境，并将任务进行分发和结果汇总等能力。
- **数据库适配层**：使用 OpenCrawler 获取的数据，除存储到文件系统的数据外，很多数据需要持久化到对应的数据库中以备使用。数据库适配层提供了一些常见数据库的存储过程管理，如 ClickHouse、Neo4j 等。
- **定时任务管理器**：OpenCrawler 提供了定时任务管理器，可通过 cron 格式配置的方式对离线任务进行定时调度，以实现数据的持续收集与更新。

在采集协作日志数据时，OpenDigger 使用 OpenCrawler 的大文件下载管理器并行下载 GHArchive 数据文件，然后通过多核任务管理器对文件进行 gzip 文件格式校验与 JSON 内容格式校验。对日志内容进行并行解析后，通过数据库适配层存储到 ClickHouse 与 Neo4j 数据库中。对于额外需要通过 GitHub REST API 采集到的数据，使用 HTTP 请求管理器配合 token 管理器进行持续采集与更新。

### 2.1.3 数据存储

OpenCrawler 采集的数据将用于 OpenDigger 的后续使用。考虑到 GitHub 日志数据的数据量及特性，本信息系统选择了 ClickHouse 与 Neo4j 作为联机分析型数据库（OLAP）[23]和图数据库[24-25]。

**（1）联机分析型数据库**

开源行为的日常分析任务大多为统计型任务，考虑到如下几个 GitHub 日志数据的特征：

- 日志数据的数据量巨大。传统的 MySQL、Oracle

等关系型数据库完成全域日志数据的聚合分析时性能较弱。

- 日志数据的特点为严格持续新增。对历史数据仅有查询操作，不会进行删改操作。
- 日志数据中不同类型事件的详细数据是异构的。在特定数据表结构设计下，单条日志的存储只会使用部分数据列，存储相对稀疏。

基于上述特点，本文选择使用 ClickHouse 作为日志数据的存储数据库，ClickHouse 数据库是一款开源的用于联机分析处理的列式数据库管理系统，其性能卓越，擅长超大规模数据的交互式分析。同时列式存储的技术方案，使其在查询时只会读取当前查询语句使用到的数据列，更适合稀疏存储的宽表数据。

本文将 GitHub 日志数据的不同事件类型进行拆分细化，合并通用列，删除了部分重复和冗余的数据，最终给出一个 138 列的数据库表结构用于日志数据存储，关于表结构的字段类型说明见 OpenDigger 系统的项目仓库[3]。

**（2）Neo4j 图数据库[25]**

除日常统计分析型任务外，OpenDigger 提供图分析能力，并提供了独特的图分析模型的实现（详见 3.2 节）。由于 GitHub 日志数据内容为某开发者在某时间点对某仓库进行了某操作，该数据天然可被处理为图数据，形成一个基于 GitHub 全域仓库和开发者的协作关系网络。对这类数据，较适合使用图数据来进行存储和分析，因此我们选择了 Neo4j 图数据库作为底层存储与计算层。

根据 DBEngine 的排名[4]，在图数据库领域，Neo4j 遥遥领先，是目前全球使用最广泛的图数据库软件。Neo4j 提供了开源的社区版和付费的商业版，社区版仅支持单机部署，商业版支持分布式部署。Neo4j 社区版支持十万亿节点和百万亿边的图存储和分析能力，可以较好的支撑目前 GitHub 的协作日志数据量。

从性能、硬件成本、分析需求等角度考虑，本文在构建 GitHub 日志数据网络时仅使用了常用的部分事件类型和数据。图数据库的元路径如图 4 所示，截至 2023 年 6 月，OpenDigger 使用协作日志数据在图数据库构建的 GitHub 全域协作网络图的规模约为 8 亿节点，19 亿条边。

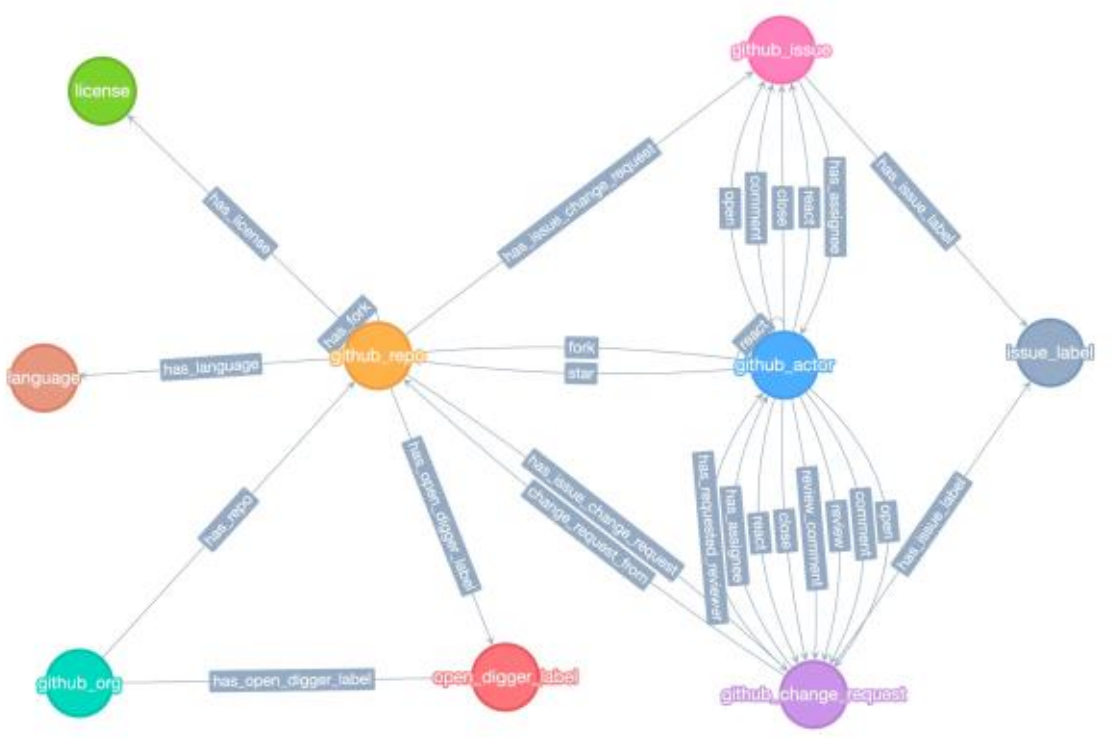

Fig. 4 The Meta Path of the Graph Database

图 4 图数据库的元路径

## 2.2 数据挖掘模块

### 2.2.1 指标体系

OpenDigger 系统旨在为开源项目提供数据分析和洞察服务。在指标实现方面，可以借鉴目前已有的一些工作，如 Linux 基金会下属的 CHAOSS 开源项目[26]。该项目长期致力于研究开源社区的健康度度量，专门面向开源项目健康度的指标定义、实现与发布。在该项目于 2022 年 4 月发布的指标体系中，包含了五个工作组共计 75 个指标项。其中通用工作组和演化工作组的指标较适合通过日常研发数据进行统计展示，而多样化与包容性工作组、风险工作组和价值工作组包含了大量的定性指标，不易通过客观数据进行统计展示。

OpenDigger 系统结合对 GitHub 日志数据与 CHAOSS 指标体系的深入分析，已实现了其中 17 个易于量化的数据指标，如表 1 所示。

### 2.2.2 分析模型

除 CHAOSS 定义的指标外，OpenDigger 也根据 GitHub 日志数据持续定义、构建与实现可供分析的指标模型。相较而言，CHAOSS 指标体系中的指标会细致的关注项目在各个方面可以采集到的客观统计数据，OpenDigger 所内置的分析模型则是基于客观数据维度进行建模后的信息模型，可以落地于生产环境中用于辅助业务决策。本文将详细介绍的两个分析模型分别是：

[3] https://github.com/X-lab2017/open-digger
[4] https://db-engines.com/en/

**Table 1　The CHAOSS Metrics Implemented by OpenDigger**

**表 1 OpenDigger 实现的 CHAOSS 指标**

| 工作组 | 指标 | 描述 |
|---|---|---|
| 通用 | 活跃的日期和时间 | 贡献者活跃的日期和时间戳是什么时候？ |
| 通用 | 技术分支 | 在代码托管平台一个开源项目有多少技术分支？ |
| 通用 | 新贡献者 | 有多少贡献者为既定项目做出第一次贡献？他们是谁？ |
| 通用 | 不活跃贡献者 | 在特定时间段内有多少贡献者处于不活跃状态？ |
| 风险 | 巴士系数 | 如果最活跃的人离开，项目会产生多大风险？ |
| 演化 | 新议题 | 在一定时期内创建了多少个新议题？ |
| 演化 | 关闭的议题 | 在一定时期内有多少关闭的议题？ |
| 演化 | 议题响应时间 | 从议题创建到另一贡献者对议题做出响应经过了多长时间？ |
| 演化 | 议题解决时长 | 解决议题花费了多长时间？ |
| 演化 | 议题时长 | 未解决的议题有多长时间处于未解决状态？ |
| 演化 | 代码变更行数 | 在一段时间内对源代码的所有更改中，所触及的行数之和（增加的行数加上移除的行数）是多少？ |
| 演化 | 变更请求 | 在一段时间内发生了哪些新的变更源代码的请求？ |
| 演化 | 接受的变更请求 | 在一次代码变更中存在多少接受的变更请求？ |
| 演化 | 变更请求评审 | 在一定时期内有多少针对变更请求的评审？ |
| 演化 | 变更请求响应时间 | 从变更请求创建到另一贡献者对变更请求做出响应经过了多长时间？ |
| 演化 | 变更请求解决时长 | 接受或关闭变更请求花费了多长时间？ |
| 演化 | 变更请求时长 | 从代码变更请求开始到接受或关闭需要多长时间？ |

- **活跃度模型**：一种基于多种事件类型加权统计下的开发者与项目活跃度评估模型。该模型在开发者侧的实现逻辑将在第三章中详细展开。
- **影响力模型**：一种基于异质图网络降维算法的开发者与项目影响力评估模型。该模型在项目侧的实现逻辑将在第三章中详细展开。

### 2.2.3 标签数据

为提供更加灵活的查询和聚合逻辑，OpenDigger 提供了如技术领域、地域、企业等不在 GitHub 日志数据中的标签数据，可用于在指标查询中提供项目、开发者筛选范围和聚合方式。目前 OpenDigger 中支持的标签数据类型包含：企业、基金会、社区、项目、发起方国家、技术领域、项目类型等。截至 2023 年 6 月，OpenDigger 包含标签数据共计 275 条，其中包含企业数据 137 条、社区项目数据 38 条，技术领域数据 78 条，项目类型数据 5 条，项目归属国家数据 11 条，基金会数据 4 条。涉及 GitHub 组织数量 413 个，涉及仓库数 89427 个。

#### （1）标签数据构建

OpenDigger 提供了一种基于规则的标签维护体系，通过文件夹关系提供标签层级结构。每个标签数据包含如下字段：

- *id*：该标签数据的唯一标识 ID；
- *name*：该标签的人类可读名称；
- *type*：该标签的标签类型；
- *data*：标签包含的数据信息，包括 GitHub 组织 ID、仓库 ID、开发者 ID 等，也可通过 label 字段直接引入其他标签数据。

#### （2）标签数据运算

OpenDigger 提供的标签数据运算支持基于标签数据 ID 和类型的协同运算。例如可以通过标签求交获得由中国发起的捐入到基金会的项目，其实现为 *labelIntersect(":regions/CN", "Foundation")*，其中中国项目是由标签数据 ID 直接指定的，而基金会则使用了基金会类型，则会引入所有的各个基金会的标签数据。而如果需要获得由中国发起的 Linux 基金会项目，则其实现为 *labelIntersect(":regions/CN", ":foundations/linux_foundation")*。这里国家与基金会均使用了标签数据 ID 进行标签数据运算。

#### （3）自定义标签数据

在实际业务应用场景中，OpenDigger 的用户可能需要使用不在 OpenDigger 中提供的公开标签数据来进行数据筛选与聚合，在该类场景下，OpenDigger

支持在指标查询中注入自定义标签数据。用户可以按照标签数据的数据格式规范创建自定义标签数据，并通过 *injectLabelData* 字段注入。

## 2.3 信息服务模块

### 2.3.1 可视化数据服务

OpenDigger 在数据库、标签数据与指标体系之上，提供了一系列可供业务侧使用的数据服务。图 5 展示了 OpenDigger 为任意活跃项目开放提供的各类指标与分析模型下的可视化数据服务，用户在 CodePen 环境中注入自己的开源项目名称即可一键拉起对应指标或分析模型下的可视化分析。

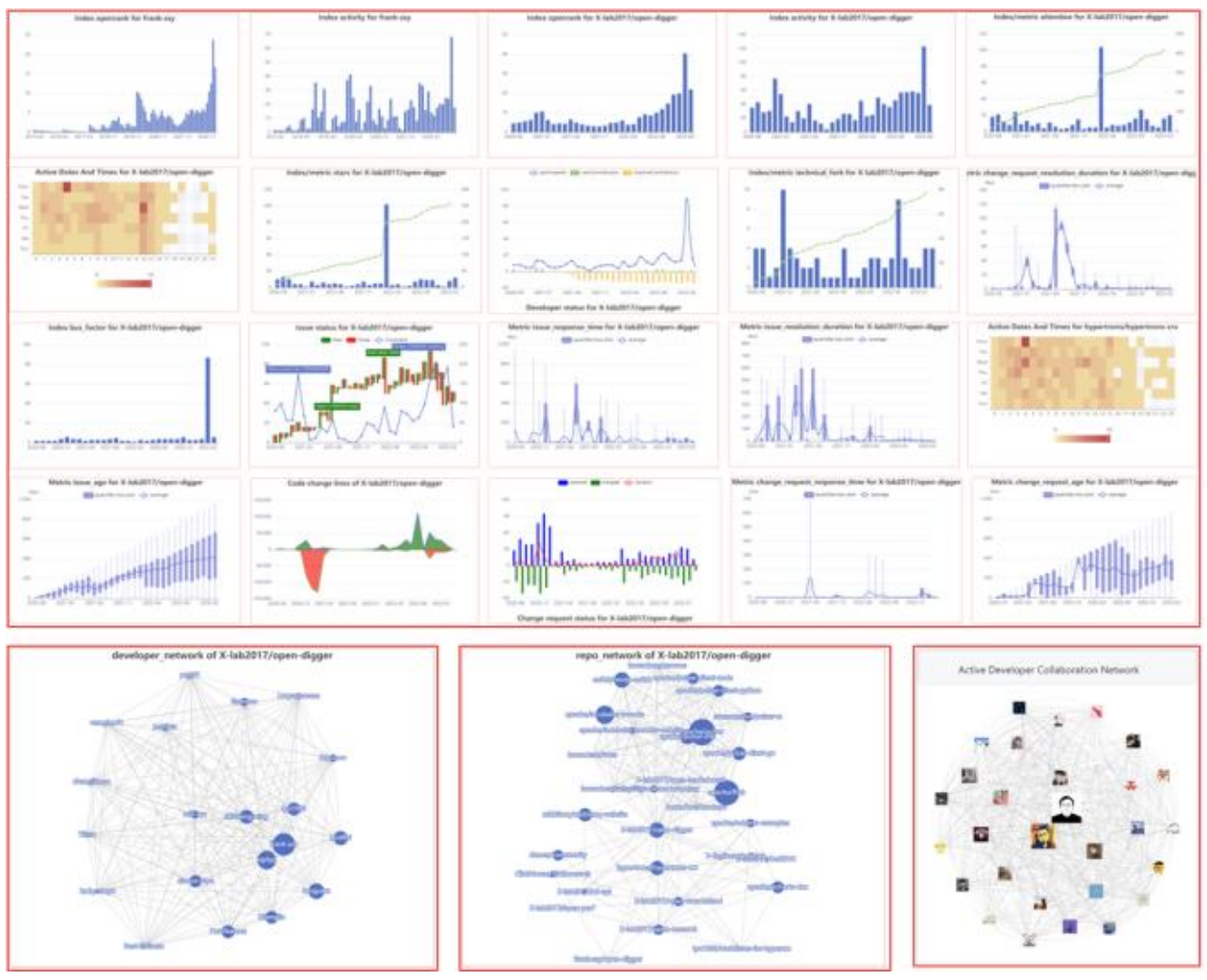

Fig. 5 Visualizations of the Data Service

图 5 数据服务的可视化展示

### 2.3.2 数据开放接口

OpenDigger 作为一个信息服务平台，提供了在 ClickHouse 数据库与应用层之间的指标和分析模型的实现逻辑，以便应用层可以直接根据需求进行调用，而无需自己实现复杂的 SQL 语句进行查询。所有 OpenDigger 所实现的指标与分析模型，均提供了指标获取接口，该接口支持表 2 所示参数。基于这些配置项，对于任意 OpenDigger 实现的指标与模型，上层可以通过配置实现对于任意范围的组织、仓库、开发者、标签交并集合进行指标查询，同时支持最小以月为粒度的时段拆分、自定义聚合方式、自定义标签数据注入、自定义排序和返回数量限制等，给予应用层足够强的可定制性，同时屏蔽了构建 SQL 的复杂度。

**Table 2 Data Interface Parameters**

**表 2 数据接口参数**

| 参数名称 | 描述 |
| --- | --- |
| startYear | 本次数据查询的起始年份，默认为 2015 |
| endYear | 本次数据查询的终止年份，默认为当前年份 |
| startMonth | 本次数据查询的起始月份，将应用于起始年份，默认为 1 |
| endMonth | 本次数据查询的终止月份，将应用于终止年份，默认为当前月份 |
| repoIds | 本次数据查询的仓库 ID 集合 |
| repoNames | 本次数据查询的仓库名称集合 |
| orgIds | 本次数据查询的组织 ID 集合 |
| orgNames | 本次数据查询的组织名称集合 |
| userIds | 本次数据查询的用户 ID 集合 |
| userNames | 本次数据查询的用户名称集合 |
| labelUnion | 标签数据并集 |
| labelIntersect | 标签数据交集 |
| order | 排序方式，可选降序排序（“DESC”）或升序排序（“ASC”），默认为“ASC” |
| orderOption | 排序选项，当存在多个时间段时，可选使用最后一个时间段排序（“latest”）或使用所有时间段之和排序（“all”），默认为“latest” |
| limit | 返回数据行数限制，默认为 10 |
| limitOption | 返回数据行数限制选项，当存在多个时间段时，可选按总返回行数限制（“all”）或按每个时间段限制（“each”），默认为“all” |
| groupBy | 仓库指标数据的聚合粒度，可选为按组织聚合（“org”）或按任意标签聚合，默认为空，即不聚合，按照仓库粒度返回 |
| groupTimeRange | 当返回多个时间段时，时间聚合的粒度，支持按年度聚合（“year”）、按季度聚合（“quarter”）和按月度聚合（“month”），默认为空，即将所有时间段聚合为一个数据点 |
| precision | 返回浮点数类型指标的精度，默认为 2，即小数点后保留 2 位 |
| injectLabelData | 可根据规范传入不在 OpenDigger 标签数据中的自定义标签数据，可用于标签数据交并运算和标签聚合 |
| options | 其他与特定指标相关的选项 |

### 2.3.3 交互式分析工具

OpenDigger 还向业务层提供了多种可交互使用的数据服务工具。分别为：

- **Node.js Core JupyterNotebook**：OpenDigger 内置了一套可用于 Node.js Core JupterNotebook 使用的完整 API，同时在官方提供的 Notebook 实例中内置了基于 Plotly.js 的图表可视化能力、标签数据访问、底层数据库的直接访问能力等，可用于灵活的进行各类数据分析工作。
- **Python Core JupyterNotebook**：与 Node.js Core JupyterNotebook 类似，OpenDigger 也提供了一套 Python 的 API 和 Notebook 实例，提供基于 Python 语法的交互式分析能力。
- **CLI 工具**：OpenDigger 实现了一个可在终端窗口中运行的工具，使用户可以通过输入命令来进行数据查询、过滤、排序、聚合、统计与基础的图表生成。该工具内置了已实现的指标和分析模型，提供高效、灵活和可扩展的方式来提供原始数据查询接口与数据分析服务。

此外，OpenDigger 还会针对 GitHub 全域项目和开发者，把其中较活跃的项目与开发者的所有历史指标数据，按照月为粒度生成所有已实现指标，并发布到云存储服务中，完全公开提供给下游使用。截止 2023 年 6 月，OpenDigger 每月生成约 60 万个仓库和 40 万个开发者的指标数据，指标文件共计 2400 余万。

## 3 OpenDigger 的分析模型与数据挖掘

通过 OpenDigger 系统所集成的全域开源协作数据，我们可以构建业界所关心的业务指标，并进行开源生态数据的挖掘与分析。本章从统计模型建模和协作网络建模两个视角，介绍 OpenDigger 所实现的两个分析模型：开源开发者活跃度与开源项目影响力，并展示案例分析与数据洞察结果。

### 3.1 基于统计模型的开发者活跃度

#### 3.1.1 分析模型建模

OpenDigger 所实现的开发者活跃度评价基于开发者在开源软件仓库中的活动，是一种通过统计开发者协作行为数据并加权求和的方式来计算开发者活跃度的计算方法。对开发者活跃度的定义，我们沿用 Xia 等人发表的工作[27]。

GitHub 以大规模协同开发与快速交付为主要特性，以 commit 为核心的 Git 事件活动并不能反映开发者的行为全貌与开源协作的特性[28]。因此，本文提出的活跃度模型基于 GitHub 平台以 Issue 和 Pull Request（PR）为协作模式的特征，融合了五种协作行为类型，即：评论、新建 Issue、新建 PR、审查 PR 和 PR 合入。具体计算公式如下：

$$A_d = \sum w_i c_i \tag{1}$$

其中，$A_d$ 为开发者活跃度，$c_i$ 为上述五种行为事件由该开发者触发的发生次数，$w_i$ 为该行为的加权比例。对五种协作行为的细节描述如表 3 所示。基于 GitHub 所采取的 pull-based 的贡献协作模型[29]，这五种行为类型涵盖了在 GitHub 上进行软件开发与贡献时的大多数协作情境。我们没有考虑 fork 和 star 行为，因为这两种行为往往是一次性的，且不涉及协作；而像 push、release 等其他行为也存在类似的问题：（1）没有协作；（2）需要特殊权限。对于行为权重的配置，本文参照了前人的工作[27]。在实际的业务场景下进行社区贡献者活跃度分析时，可以由社区负责人主观进行个性化配置。

**Table 3 The Five Collaboration Behaviors**

**表 3 五种协作行为描述**

| 行为类型 | 事件类型 | 描述 | 权重 |
|---|---|---|---|
| Comment | IssusCommentEvent | 在 Issue 或 PR 下发表一条评论 | 1 |
| OpenIsssue | IssuesEvent | 新建一个议题 | 2 |
| OpenPR | PullRequestEvent | 新建一个代码变更请求 | 3 |
| ReviewPR | PullRequestReviewCommentEvent | 发起一次审查代码变更请求 | 4 |
| PRMerged | PullRequestEvent | 代码变更请求被合入 | 2 |

#### 3.1.2 数据洞察

本节使用 2019 年全年 GitHub 日志进行案例分析，总日志条数约 5.46 亿条，相较 2018 年的 4.21 亿条增长约 29.7%。在上述开发者活跃度的定义下，统计得到 2019 年总活跃开发者数量约 360W，相较 2018 年的活跃开发者数增长约 18.8%（2018 年约 303W）。根据上述活跃度定义，我们对 2019 年全年活跃开发者进行了活跃度统计与排名，得到世界活跃度 Top 10 开发者账号与活跃度定义中的五种行为事件数量分布如表 4 所示。

**Table 4　Top10 GitHub Active Developers in 2019**

**表 4　2019 年 GitHub 开发者活跃度 Top10**

| 排名 | 仓库 | 活跃度 | Comment | OpenIssue | OpenPR | ReviewPRc | PRMerged |
|---|---|---|---|---|---|---|---|
| 1 | dependabot[bot] | 3641110 | 272213 | 0 | 3006606 | 0 | 374345 |
| 2 | direwolf-github | 589065 | 0 | 0 | 340105 | 0 | 0 |
| 3 | dependabot-preview[bot] | 503923 | 808143 | 19595 | 1675424 | 0 | 790491 |
| 4 | pull[bot] | 281141 | 0 | 0 | 1127083 | 0 | 1124472 |
| 5 | github-learning-lab[bot] | 262617 | 372192 | 142165 | 43742 | 12864 | 33362 |
| 6 | renovate[bot] | 200693 | 37663 | 1807 | 574740 | 0 | 455955 |
| 7 | greenkeeper[bot] | 147530 | 825189 | 56071 | 161884 | 0 | 81707 |
| 8 | imgbot[bot] | 105982 | 0 | 0 | 58822 | 0 | 11459 |
| 9 | autotester-one | 88989 | 0 | 89029 | 0 | 0 | 0 |
| 10 | snyk-bot | 88096 | 1 | 0 | 126804 | 0 | 10589 |

从数据洞察结果来看，世界活跃度 Top 10 的开发者账号均为机器人账号，其中 7 个账号为 GitHub App。从不同行为事件类型的数量分布上可以看出，PR 的自动化开启与合入是最常用到的行为，其次是评论。从中可以观察与推测开发者的自动化协同习惯：开发者主要使用自动化仓库管理工具进行依赖更新、自动同步上游、GitHub 学习、漏洞检测等功能。

## 3.2 基于异质网络降维的项目影响力

GitHub 承担了协作与社交的功能，其上的行为数据不仅可以进行线性的统计，而且可以构成一个一般意义上的社交网络。因此，在实现 CHAOSS 中的统计型指标和开发者活跃度统计模型之外，OpenDigger 还探索基于 GitHub 全域行为数据的协作关系网络，并利用一种加权 PageRank 算法计算项目的影响力。

### 3.2.1 分析模型建模

项目影响力的计算基于异质信息网络的构建和降维[4]，构造网络的方式是构建一个开发者和项目的二部图（Bipartite Graph）[30]。开发者和项目是两类节点，而所有的边都在这两类节点之间，即不会出现开发者之间的边或项目之间的边。图 6 是一个简单的二部图，其中 *pa* 和 *pb* 是两个项目，*da* 和 *db* 是两个开发者。*da* 曾在 *pa* 内开过一条 Issue，而 *db* 既在 *pa* 内有过 Issue 或 PR 下的评论，又向在 *pb* 内有过 PR 的发起和评审。直接对于如图所示的异构图进行分析是较为复杂的，传统的图算法分析大多基于同质信息网络，因此可以对这个网络图进行降维，使其变为一个同质信息网络。

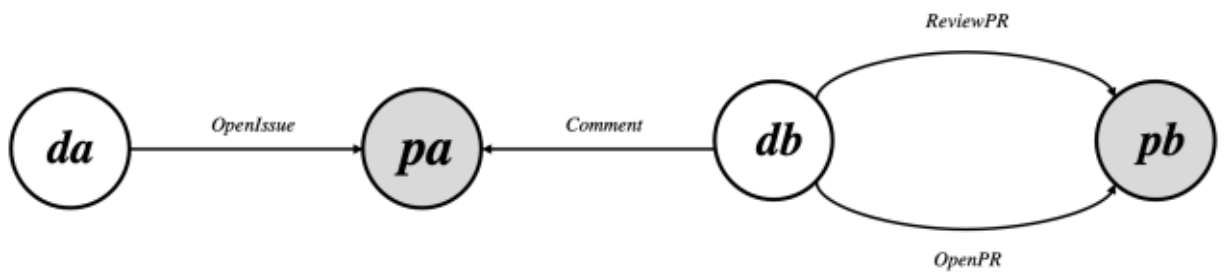

Fig. 6 Heterogeneous Collaboration Network

图 6　异质协作网络

**（1）关系类型降维**

在图 6 所示的异质图中，依然使用 3.1 节所介绍的活跃度模型中的五种协作行为类型构造完整的协作网络。则对于这个网络，可以使用活跃度模型进行降维。具体为：对于开发者和项目之间的边，在异质结构下包含各种不同的类型，例如 *OpenIssue*、*Comment*、*ReviewPR* 等，通过开发者活跃度计算公式，可以将所有的不同类型的边压缩成一种类型，即 *Active*，也就是活跃。这条边可以包含的属性是该开发者在该项目上的活跃度。根据表 3 中的行为类型权重，可得 *da* 在 *pa* 上的活跃度为 2，在 *pb* 上不产生活跃。而 *db* 在 *pa* 上的活跃度为 1、在 *pb* 上的活跃度为 7（打开一条 PR 的活跃值为 3，审查一条 PR 的活跃值为 4）。如图 7 所示。

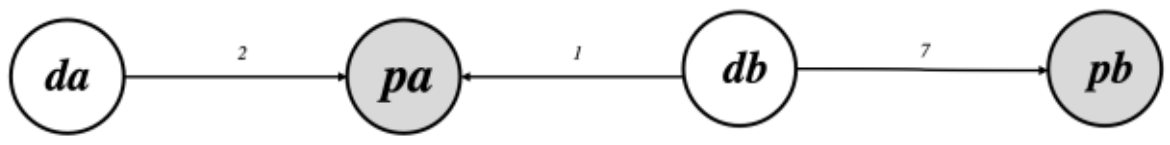

Fig. 7　Dimensionality Reduction of Relationship

图 7　关系类型降维

**（2）节点类型降维**

由于希望计算的是项目的影响力，所以在进行节点类型降维时应想办法消除开发者节点。这里将项目的关联关系定义为具有相同的活跃开发者，即若开发者同时在两个项目内活跃过，则该开发者就关联了这

两个项目。对这种关联关系的量化，我们使用开发者在两个项目上的活跃度的调和平均作为该开发者对两个项目关联度的贡献，如公式所示，

$$R_{papb} = \sum_{dev} \frac{A_{da}A_{db}}{A_{da}+A_{db}} \qquad (2)$$

*pa*、*pb* 两个项目的关联度，为所有在两个项目上同时活跃过的开发者在两项目上的活跃度调和平均和。由于 *da* 只在 *pa* 项目中活跃，所以不对两个项目的关联度有贡献。而 *db* 在两个项目中均有活跃，所以可以对两个项目关联度有贡献，且贡献值是 *7/(1+7) = 0.875*。通过这种方式，可以将开发者从图中移除，构成一个仅包含项目和它们之间关联度的同质信息网络图，如图 8 所示。

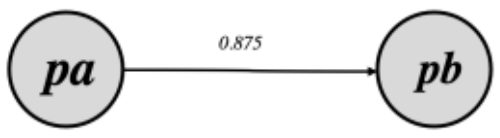

Fig. 8 Dimensionality Reduction of Node

图 8 节点类型降维

**（3）影响力计算**

通过上述方法和 OpenDigger 提供的全域数据，可以构建在一段特定时间内的全域项目的同质网络图。对于图的分析，采用加权 PageRank 算法（Weighted PageRank，后称 WPR）[31]计算节点的中心度，该中心度的值为计算得到的项目影响力指标。

PageRank 算法是 Google 最著名的网页排序算法[32]。经典的 PageRank 算法是进行有向无权图计算的一种算法，边是网页之间的链入关系。而在本文中，各节点之间是无向的，也就是: $R_{ab} = R_{ba}$。其次是带权图，即每两个项目之间的关联度存在一个量化指标，这个指标确定了两个项目之间的关联紧密程度，故经典 PageRank 算法并不能直接适用。于是我们引入 WPR 算法[31]，WPR 算法在经典的 PageRank 算法中引入边权，即迭代计算时，节点中心度的值不再平均分配到相邻节点，而是通过边权来确定分配比例。

另外，由于 WPR 也是计算有向图中心度，所以在建立边时，将同时建立两个项目指向彼此一条有向边，而且边权相等，均为上述计算出的两个项目的关联度。本文直接对构建出的网络进行了全域计算。但由于 GitHub 上存在大量的自动化机器人账号，可能对项目关联度造成影响，故在建图后过滤了当年活跃数量大于 200 个项目的开发者账号的数据。

### 3.2.2 数据洞察

**（1）整体结果**

本节依旧针对 2019 年全年 5.46 亿条 GitHub 日志进行案例分析。2019 年 GitHub 活跃仓库数共计 3972.3W 个，剔除 173 个活跃数量大于 200 个的开发者账号与其相关数据后，剩余活跃仓库数量共计 870249 个。最终构成的全域项目关联网络中节点数量为 870249 个，边数量为 4320.3W 条。首先对于该图进行连通性分析，总共包含 30870 个强连通分量，其中最大的连通分量包含 786659 个项目，即 GitHub 中 90.4% 的项目在上述关联度定义下是连通的，剩余 83590 个项目分成了 30869 个独立的分图，其中最大的连通分量包含 156 个项目，其他大部分项目为单项目节点，即与其他任何项目都没有产生关联。

我们对其中项目数大于 50 的连通分量进行进一步观察，发现大部分项目为测试项目，可能由一个账号批量创建的。而另外的群落则表现出明显的地理隔离，其中较大的一些隔离群落分别来自日本、俄罗斯、法国、乌克兰、白俄罗斯，未发现有中国开发者群聚而出现地理隔离的情况。说明语言障碍很可能是导致发生开源社区隔离的一种重要原因。

**（2）项目排名对比**

该项目影响力计算结果可以与 GitHub 官方发布的 Octoverse 报告[5]进行对比。结合具体案例，可以说明基于协作网络计算方式的有效性。Octoverse 2019 年度的 Top10 项目使用项目贡献者数量作为评价维度。表 5 展示了使用影响力排名和贡献者数排名的项目 Top10 结果对比，横线加粗部分为同时出现在榜单上的项目，重合度达到 60%。且影响力排名高的项目也都耳熟能详，符合大众认知。而在影响力排名 Top20 中，除 first-contributions 项目外，Octoverse Top10 的项目中有 9 个项目均在榜上。

**（3）案例分析**

***microsoft/vscode*** 项目在所有排名中都名列第一，且影响力计算结果一骑绝尘，是第二名 flutter/flutter 的将近两倍。表 6 展示了与 vscode 关联度最高的 10 个项目的关联度与自身影响力。在这 10 个项目中，有多达 4 个项目与 vscode 一同处在全球影响力 Top 10，包含了编 vscode 的语言项目 TypeScript、与 TypeScript 高度相关，给 Node.js 类库添加类型说明的 DefinitelyTyped、Google 开源的跨平台应用开发框架 flutter 和基于 React 的快速开发框架 gatsby。而该项目与自己的开发语言之间的关联度高达近 800。从网络关系中可以看到 vscode 与一众顶级项目一起构成了一个庞大的开源社群，他们之间关系密切而又相互促进，吸引着世界上最优秀的开发者共同参与其中协同与贡献。

[5] https://octoverse.github.com/2019/

***firstcontributions/first-contributions*** 项目因贡献者众多，在 Octoverse 2019 的报告中排名全球第四，而在项目影响力计算下，全球排名是 532，影响力值仅有 49。在不同的算法下如此大的差距与该项目本身的特性有很大关系。作为一个 GitHub 上的学习型项目，该项目中包含了数十种语言的如何提交 GitHub 第一个 PR 的教程。大量初学者会选择用该项目来学习提交 PR，故在项目上产生了大量的贡献者。所以当 GitHub 以活跃贡献者数量排名时，该项目排行全球第四。表 7 是与该项目关联度最高的 10 个项目，以及他们的关联度和这些项目自身的影响力值。可以看到，除 gatsby 外，其他项目的影响力均不高。更关键的是，该项目与这些项目之间的关联度也很低，即便是关联度最高的 freeCodeCamp 项目关联度也只有 28。而一般强关联项目的关联度均可到达到数百甚至上千，例如 kubernetes/kubernetes 与 kubernetes/enchancements 两个仓库的关联度高达 1815。造成这种现象的原因可能包含两个：

1）由于这是一个学习型项目，活跃在项目上的开发者均为 GitHub 的初学者。他们并不会在短时间内成为其他顶级项目的核心开发者，所以该项目很难与各种顶级项目之间产生关联。

2）同样由于这是一个学习型项目，大部分开发者与该项目的关联都是一次性的。即使少量的初学者后来成为了顶级项目的开发者，由于与该项目的关联较弱，也不会导致该项目与顶级项目之间产生强关联。例如 gatsbyjs/gatsby 虽然自身排名很高，但与该项目的关联度只有 11，所以很难对该项目的排名产生质的影响。 first-contributions 项目在贡献者数排名与影响力排名的对比很好地诠释了影响力指标基于协作关联网络进行计算的优势，即开发者与项目、以及项目之间的通过开发者的行为关联了起来。这些关联被很好的量化，最终产生了一个价值网络。

**Table 5 Results Comparison of Project Influence Ranking and Octoverse Ranking**

**表 5 项目影响力排名与 Octoverse 排名结果对比**

| 影响力排名 | 仓库 | 影响力 | 贡献者数排名 | 仓库 |
|---|---|---|---|---|
| 1 | **microsoft/vscode** | 1135 | 1 | **microsoft/vscode** |
| 2 | **flutter/flutter** | 645 | 2 | MicrosoftDocs/azure-docs |
| 3 | **kubernetes/kubernetes** | 624 | 3 | **flutter/flutter** |
| 4 | **DefinitelyTyped/DefinitelyTyped** | 564 | 4 | firstcontributions/first-contributions |
| 5 | microsoft/Typescript | 544 | 5 | **tensorflow/tensorflow** |
| 6 | **tensorflow/tensorflow** | 535 | 6 | **facebook/react-native** |
| 7 | gatsbyjs/gatsby | 504 | 7 | **kubernetes/kubernetes** |
| 8 | golang/go | 448 | 8 | **DefinitelyTyped/DefinitelyTyped** |
| 9 | rust-lang/rust | 448 | 9 | ansible/ansible |
| 10 | **facebook/react-native** | 426 | 10 | home-assistant/home-assistant |

**Table 6 Top 10 Most Relaed Projects of vscode**

**表 6 vscode 关联度 Top10 项目**

| 仓库 | 关联度 | 影响力 |
|---|---|---|
| microsoft/TypeScript | 799 | 544 |
| microsoft/vscode-remote-release | 594 | 162 |
| microsoft/vscode-python | 458 | 157 |
| DefinitelyTyped/DefinitelyTyped | 410 | 564 |
| Microsoft/vscode-cpptools | 360 | 102 |
| microsoft/Terminal | 323 | 243 |
| electron/electron | 255 | 256 |
| flutter/flutter | 236 | 645 |
| microsoft/vscode-docs | 227 | 40 |
| gatsbyjs/gatsby | 220 | 504 |

**Table 7 Top 10 Most Related Projects of first-contributions**

**表 7 first-contributions 关联度 Top10 项目**

| 仓库 | 关联度 | 影响力 |
|---|---|---|
| freeCodeCamp/freeCodeCamp | 28 | 77 |
| pandas-dev/pandas | 15 | 195 |
| firstcontributions/firstcontributions.github.io | 15 | 4 |
| gatsbyjs/gatsby | 11 | 504 |
| publiclab/plots2 | 10 | 52 |
| scikit-learn/scikit-learn | 9 | 138 |
| ows-ali/Hacktoberfest | 9 | 3 |
| systers/mentorship-backend | 9 | 5 |
| Ishaan28malik/Hacktoberfest2019 | 8 | 5 |
| danthareja/contribute-to-open-source | 8 | 4 |

其中，有价值的项目和开发者被关联在一起，并可以直接通过算法进行计算和排名，在关联关系的视角下，提供了非常有启发性的生态洞察。

### 3.3 讨论

本章介绍了 OpenDigger 信息系统中提供的两个分析模型：活跃度和影响力。对于这两个指标的实现存在不同的理解视角和计算方式。两类指标的底层数学模型不同，但计算秉承的原则都是基于开发者的协作行为，而不关注代码本身的内容，践行 GitHub“社交编程”（social coding）[33]的理念。在开发者活跃度计算中，选用哪些行为，这些行为的权重设置以及是否包含例如边际收益递减[34]等经济学相关的内容，都可以在未来进一步探讨。

而项目影响力基于活跃度计算公式构建，但同时又规避了很多活跃度会带来的问题，并且利用了开源生态的全域数据所蕴含的一些重要的关联信息做进一步的数据挖掘。典型的如一些自动化项目由于特定行为的数量极大，活跃度值会异常的高，但由于其自动化账号没有与其他项目产生关联，这种异常的高度活跃在网络构建时就被消除，因此协作网络的优势体现在对刷分行为的免疫性，即对自己项目的高活跃的刷分无法带动自己项目的影响力。

此外，在网络结构的关联关系特性下，活跃度计算中对权重设置所造成的计算结果在该计算模型下也会变小。只要依旧满足泛化的价值判断，即代码贡献大于代码评审大于一般问题讨论，则活跃度中权重的变化几乎不会影响到影响力的排名。因为影响力中活跃度虽然是一个基础数据，但协作网络的结构信息使得整个算法具有了更好的稳定性和鲁棒性。

## 4 OpenDigger 信息服务的实际应用

OpenDigger 作为一个开源生态数据挖掘与数据服务的数据基础类信息系统，被广泛应用到各类产业界与学术界的真实场景中。图 9 展示了 OpenDigger 所支撑的典型业务场景和实际应用，包括社区生态端、行业端、企业端、以及高校端。

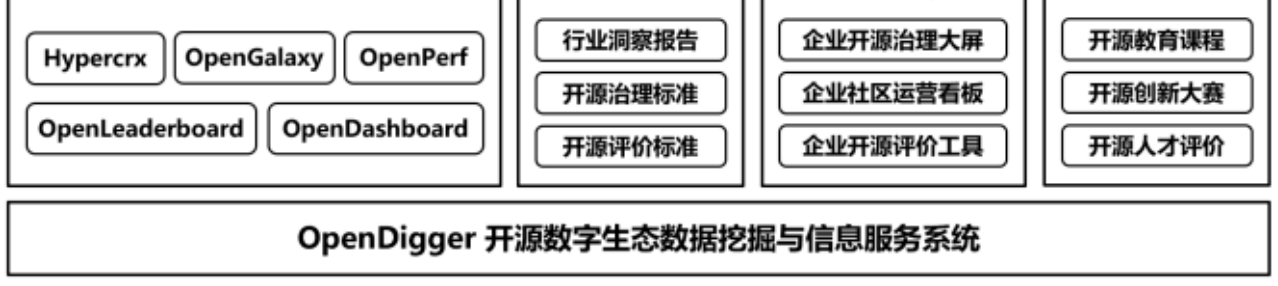

Fig. 9 Scenarios and Applications Supported by OpenDigger

图 9 OpenDigger 支持的业务场景和实际应用

在行业端，OpenDigger 提供了行业 BI 分析能力，《中国十年开源洞察报告》使用 OpenDigger 所提供的近十年开发者行为日志数据，从开源整体发展趋势、活跃开发者行为、开发者最爱的开发者语言、活跃开源项目、活跃厂商等多个维度系统展示了近十年中国开源的演进[6]；《GitHub 2020 数字洞察报告》是基于 OpenDigger，联合多家科研机构与开源社区所共同完成的一个反映全球开源现状与趋势的一个报告项目[7]。此外，近年来的《中国开源年度报告》（开源社出品）、《中国开源发展蓝皮书》（中国开源软件推进联盟出品）、《开源生态白皮书》（信通院出品）等，均采用了 OpenDigger 的信息服务作为数据源。

在高校端，OpenDigger 已经广泛的应用于各类比赛、课程、报告和论文的数据生产需求中，为这些业务场景提供数据服务。华东师范大学、湖北大学、西南民族大学等多所高校面向本科生与研究生的计算机通识课程和开源软件通识课程中已经使用 OpenDigger 提供的数据能力作为学生探索软件仓库挖掘与分析的教具[8]；OpenDigger 也参与到各类技术大赛中提供数据集和基准测试服务，如开放原子全球开源大赛[9]、CCF 中国软件开源创新大赛、Paddle 社区数据分析黑客松[10]等；OpenDigger 具备的数据生产能力和指标系统也孕育了各类研究问题，如开源项目推荐[35]、开发者行为模式识别[27]、开源社区健康度分析[36]等。下面再重点列举三个基于 OpenDigger 的典型应用。

### 4.1 社区生态端应用：Hypercrx 插件项目

Hypercrx 是一个使用 OpenDigger 提供的数据服务开发的 GitHub 可视化看板插件项目。GitHub 是许多人探索开源世界的第一站，为了使人们更好地洞察 GitHub 上的开源项目和开发者，通过消费 OpenDigger 的指标系统和指标数据接口，Hypercrx 利用可视化技术和浏览器插件技术在 GitHub 页面上嵌入了多个图表组件，如图 10 所示。其中，基于 GitHub 全域网络数据的项目协作网络图和开发者协作网络图直观展示了“项目-项目”、“项目-开发者”和“开发者-开发者”之间的联系，非常具有启发性。作为一个开源项目，Hypercrx 也吸引了许多用户和开发者的关注。[11]

[6] https://developer.aliyun.com/article/1044488
[7] http://oss.x-lab.info/github-insight-report-2020.pdf
[8] https://github.com/X-lab2017/oss101
[9] https://atomgit.com/x-lab/OpenSODA
[10] https://github.com/PaddlePaddle/Paddle/issues/43938
[11] https://github.com/hypertrons/hypertrons-crx

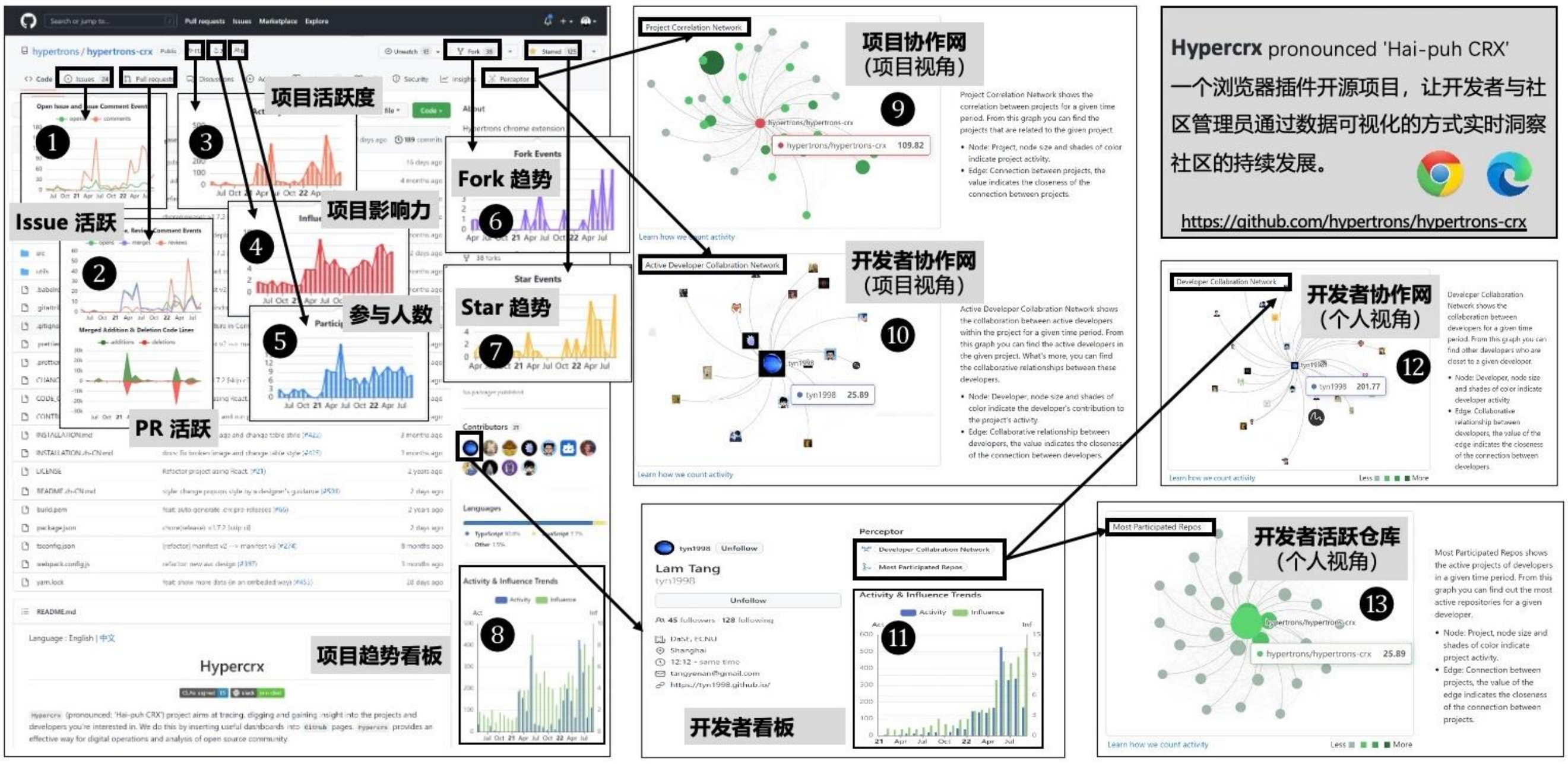

**Fig. 10** Hypercrx: GitHub Visualization Dashboard Extension Project

**图 10** GitHub 可视化看板插件项目 Hypercrx

## 4.2 行业端应用：开源治理标准与治理可视化大屏

OpenDigger 系统作为一个开源项目于 2021 年正式捐赠到中国电子技术标准化研究院旗下的木兰开源社区进行孵化，并作为国家开源治理系列标准的支撑工具。由华东师范大学所牵头研制的开源治理系列标准如图 11 所示，包括：总体框架、企业治理模型、社区治理与运营、开源项目评价模型、以及开源贡献者评价模型。

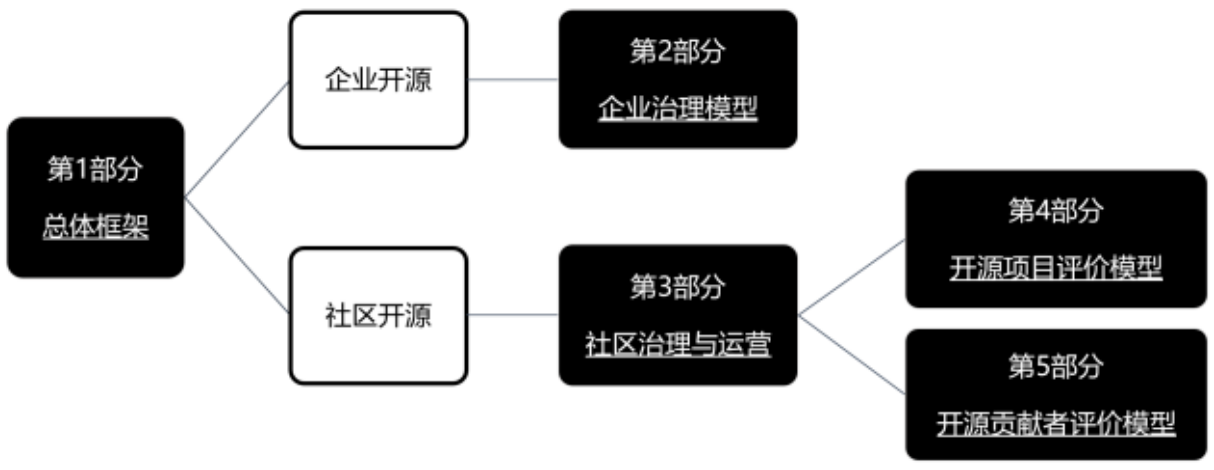

Fig. 11 Open Source Governance Standards

图 11 开源系列治理标准

这些系列标准中均采用了 OpenDigger 中的指标、算法模型、信息服务作为参考实现，并被逐步推广到业界采用。其中，木兰社区本身也基于这些标准与 OpenDigger 开发了项目孵化与治理大屏，为旗下众多项目提供信息服务，如图 12 所示。

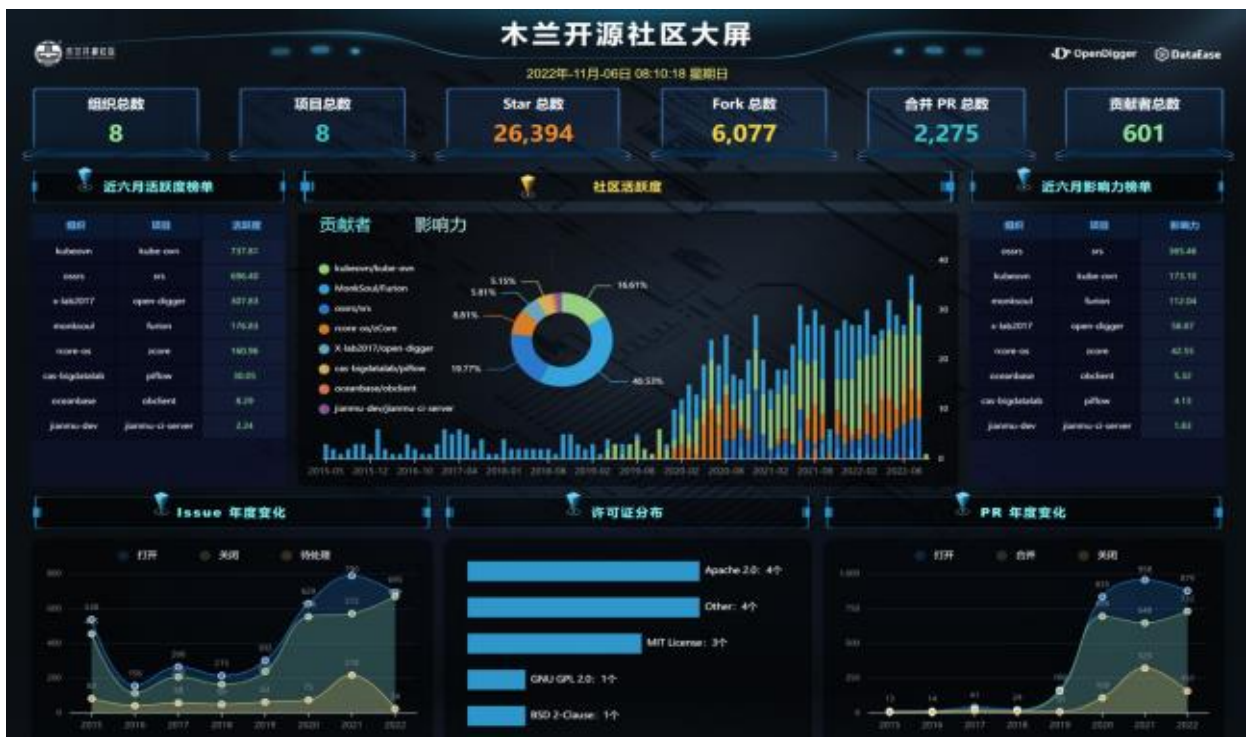

**Fig. 12** Mulan Community Open Source Governance Dashboard

**图 12** 木兰社区开源治理大屏

## 4.3 企业端应用：Nacos 开源项目运营数据大屏

OpenDigger 还服务于企业、社区、基金会、个人开发者端的开源治理大屏，为 OSPO（Open Source Program Office，开源办公室）从业者和开源项目运营负责人提供开源数字洞察能力。如图 13 是使用 OpenDigger 提供的数据服务和 DataV 技术栈实现的 alibaba/nacos 项目治理大屏[12]。该大屏展示了 nacos 项目的活跃度与影响力变化趋势、项目参与人数、issue 情况以及 PR 情况等。这些洞察有助于企业或组织及时了解自身开源项目与整体社区生态的健康状况、趋势和潜在问题，从而做出更明智的决策和优化运营策略。除图 13 所示大屏外，企业端还可以实现多项目竞品分析大屏用于获取有关同技术类型开源

[12] http://repo-data.opensource-service.cn/?r=alibaba/nacos

项目的关键洞察，从而更好地进行技术选型。利用 OpenDigger 的标签数据，特定开发者级别，项目级别，社区组织级别，以及基金会级别等一系列的数据洞察解决方案都可以一键配置和可视化实现。

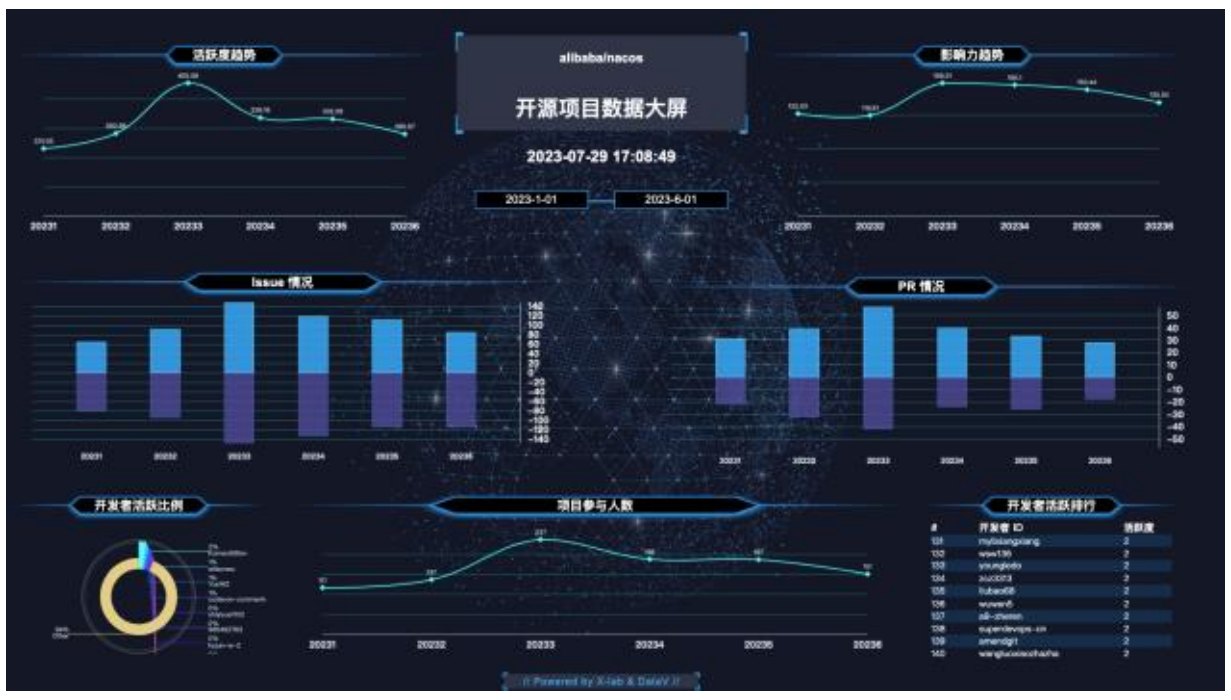

**Fig. 13** Nacos dashboard based on OpenDigger

**图 13** 基于 OpenDigger 系统的 nacos 项目数据大屏

### 4.3 高校端应用：高校开源贡献的评价与激励

高校师生参与全球开源社区的贡献是当下科技强国的重要举措，如由中央网信办、教育部正在实施的一流“网络安全学院”、“特色软件工程学院”等建设示范项目，拟在试点学校中，开展高校参与开源贡献的评价与激励工作。因此，如何科学、客观、公正地对开源贡献进行度量与评估，是有效引导广大师生、激励开源贡献者的重要抓手，是高校促进开源生态发展的“牛鼻子”工程。

目前基于 OpenDigger 数据与指标体系的高校开源贡献评价与激励工作已经在华东师范大学从 2022 年 6 月份开始做试点，将华东师范大学师生参与开源贡献的活动进行科学量化，并通过奖学金与奖教金的方式进行激励。该工作整体效果明显，能够很好的带动师生参与开源的动力，同时能够很好提高项目和社区本身的全域影响力，进而持续发展。

图 14 展示了华东师范大学 XSOSI 开源评价与激励项目在过去 12 个月的数据可视化情况。

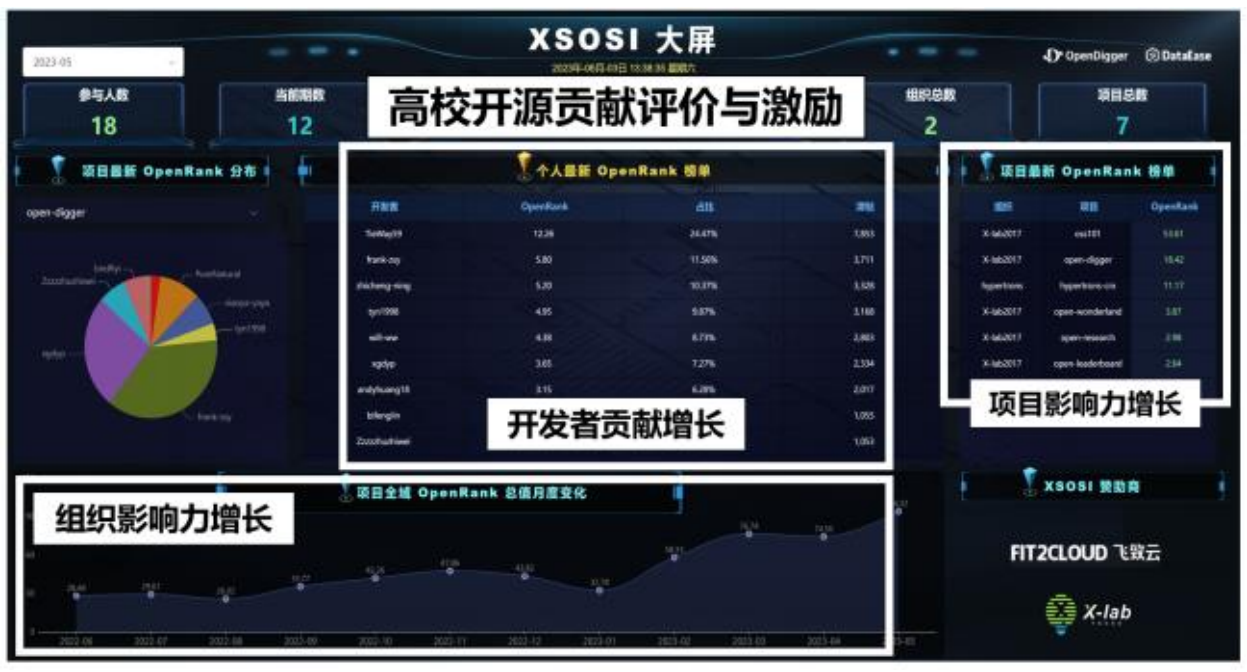

Fig. 14 Evaluation of open source contribution by OpenDigger

图 14 基于 OpenDigger 的学生开源贡献度评价

从图中可以看出，在 OpenRank 贡献度量化评价模型[37]的支持下，能带动引包括组织层面、项目层面、以及开发者个人层面的协同增长趋势，体现出 OpenDigger 推动开源生态发展的强大效应。

## 5 结论

开源软件已经成为人类数字社会的基石，是全人类共同努力的结晶。由于开源软件生态的持续发展，源源不断的开放数据集为相关研究工作带来了巨大的便利性与创新机会。OpenDigger 信息系统充分利用与挖掘开源生态中的开放数据，为研究人员、开发者、开源项目管理者与企业用户提供数据与服务支持。OpenDigger 所持续集成的开源生态领域下的业务数据使我们可以充分挖掘开源开发者的行为特征和开源项目的协作与影响力机制，分析和洞察开源生态，揭示出更深层次的信息和模式，为决策制定和业务优化提供更有力的支持。

在未来工作中，OpenDigger 信息系统将继续探索开源软件生态系统中的数据潜力，不断扩充开源生态领域下的数据服务范围，完善指标体系和分析模型，提供更多样化的数据支持，推动开源软件生态系统的进一步发展。

## 参 考 文 献

[1] 周明辉，张宇霞，谭鑫. 软件数字社会学[J]. 中国科学：信息科学,2019(11):1399-1411.

[2] Walker G H, Stanton N A, Salmon P M, et al. A review of sociotechnical systems theory: a classic concept for new command and control paradigms[J]. Theoretical issues in ergonomics science, 2008, 9(6): 479-499.

[3] Ropohl G. Philosophy of socio-technical systems[J]. Society for Philosophy and Technology Quarterly Electronic Journal, 1999, 4(3): 186-194.

[4] Chung F R K, Lu L. Complex graphs and networks[M]. American Mathematical Soc., 2006.

[5] Ma Y, Bogart C, Amreen S, et al. World of code: an infrastructure for mining the universe of open source VCS data[C]//2019 IEEE/ACM 16th International Conference on Mining Software Repositories (MSR). IEEE, 2019: 143-154.

[6] Drost-Fromm I, Tompkins R. Open Source Community Governance the Apache Way[J]. Computer, 2021, 54(4): 70-75.

[7] Yuan L, Wang H M, Yin G, et al. Mining and analyzing behavioral characteristic of developers in open source software[J]. Journal of Computers, 2010, 33(10): 1909-1918.(in Chinese)

袁霖,王怀民,尹刚,等.开源环境下开发人员行为特征挖掘与分析[J].计算机学报,2010,33(10):1909-1918.

[8] LI Cun-yan, HONG Mei. Analysis on Behavior Characteristics of Developers in Github[J]. Computer Science, 2019, 46(2): 152-158.(in

Chinese)
李存燕，洪玫. Github 中开发人员的行为特征分析[J]. 计算机科学, 2019, 46(2): 152-158.
[9] Constantino K, Souza M, Zhou S, et al. Perceptions of open-source software developers on collaborations: An interview and survey study[J]. Journal of Software: Evolution and Process, 2023, 35(5): e2393.
[10] Marlow J, Dabbish L, Herbsleb J. Impression formation in online peer production: activity traces and personal profiles in github[C]//Proceedings of the 2013 conference on Computer supported cooperative work. 2013: 117-128.
[11] Tsay J, Dabbish L, Herbsleb J. Influence of social and technical factors for evaluating contribution in GitHub[C]//Proceedings of the 36th international conference on Software engineering. 2014: 356-366.
[12] McDonald N, Goggins S. Performance and participation in open source software on github[M]//CHI'13 extended abstracts on human factors in computing systems. 2013: 139-144.
[13] Yin G, Wang T, Liu BX, Zhou MH, Yu Y, Li ZX, Ouyang JQ, Wang HM. Survey of Software Data Mining for Open Source Ecosystem. Journal of Software, 2018, 29(8): 2258-2271(in Chinese)
尹刚，王涛，刘冰珣，周明辉，余跃，李志星，欧阳建权，王怀民. 面向开源生态的软件数据挖掘技术研究综述. 软件学报，2018，29(8): 2258-2271.
[14] Samoladas I, Gousios G, Spinellis D, et al. The SQO-OSS quality model: measurement based open source software evaluation[C]//Open Source Development, Communities and Quality: IFIP 20 th World Computer Congress, Working Group 2.3 on Open Source Software, September 7-10, 2008, Milano, Italy 4. Springer US, 2008: 237-248.
[15] Bauer V, Heinemann L, Hummel B, et al. A framework for incremental quality analysis of large software systems[C]//2012 28th IEEE International Conference on Software Maintenance (ICSM). IEEE, 2012: 537-546.
[16] Zou Y, Liu C, Jin Y, et al. Assessing software quality through web comment search and analysis[C]//Safe and Secure Software Reuse: 13th International Conference on Software Reuse, ICSR 2013, Pisa, June 18-20. Proceedings 13. Springer Berlin Heidelberg, 2013: 208-223.
[17] Allamanis M, Sutton C. Why, when, and what: analyzing stack overflow questions by topic, type, and code[C]//2013 10th Working conference on mining software repositories (MSR). IEEE, 2013: 53-56.
[18] Henß S, Monperrus M, Mezini M. Semi-automatically extracting FAQs to improve accessibility of software development knowledge[C]//2012 34th International Conference on Software Engineering (ICSE). IEEE, 2012: 793-803.
[19] Wong E, Yang J, Tan L. Autocomment: Mining question and answer sites for automatic comment generation[C]//2013 28th IEEE/ACM International Conference on Automated Software Engineering (ASE). IEEE, 2013: 562-567.
[20] Dagenais B, Robillard M P. Recovering traceability links between an API and its learning resources[C]//2012 34th international conference on software engineering (icse). IEEE, 2012: 47-57.
[21] Bacchelli A, Ponzanelli L, Lanza M. Harnessing stack overflow for the ide[C]//2012 Third International Workshop on Recommendation Systems for Software Engineering (RSSE). IEEE, 2012: 26-30.
[22] Gousios G, Vasilescu B, Serebrenik A, et al. Lean GHTorrent: GitHub data on demand[C]//Proceedings of the 11th working conference on mining software repositories. 2014: 384-387.
[23] Chaudhuri S, Dayal U. An overview of data warehousing and OLAP technology[J]. ACM Sigmod record, 1997, 26(1): 65-74.
[24] Angles R. A comparison of current graph database models[C]//2012 IEEE 28th International Conference on Data Engineering Workshops. IEEE, 2012: 171-177.
[25] Miller J J. Graph database applications and concepts with Neo4j[C]//Proceedings of the southern association for information systems conference, Atlanta, GA, USA. 2013, 2324(36): 141-147.
[26] Goggins S P, Germonprez M, Lumbard K. Making open source project health transparent[J]. Computer, 2021, 54(8): 104-111.
[27] Xia X, Weng Z, Wang W, et al. Exploring activity and contributors on GitHub: Who, what, when, and where[C]//2022 29th Asia-Pacific Software Engineering Conference (APSEC). IEEE, 2022: 11-20.
[28] Young J G, Casari A, McLaughlin K, et al. Which contributions count? Analysis of attribution in open source[C]//2021 IEEE/ACM 18th International Conference on Mining Software Repositories (MSR). IEEE, 2021: 242-253.
[29] Gousios G, Pinzger M, Deursen A. An exploratory study of the pull-based software development model[C]//Proceedings of the 36th international conference on software engineering. 2014: 345-355.
[30] Asratian A S, Denley T M J, Häggkvist R. Bipartite graphs and their applications[M]. Cambridge university press, 1998.
[31] Xing W, Ghorbani A. Weighted pagerank algorithm[C]//Proceedings. Second Annual Conference on Communication Networks and Services Research, 2004. IEEE, 2004: 305-314.
[32] Page L, Brin S, Motwani R, et al. The pagerank citation ranking: Bring order to the web[R]. Technical report, stanford University, 1998.
[33] Dabbish L, Stuart C, Tsay J, et al. Social coding in GitHub: transparency and collaboration in an open software repository[C]//Proceedings of the ACM 2012 conference on computer supported cooperative work. 2012: 1277-1286.
[34] Marinazzo D, Wu G, Pellicoro M, et al. Information flow in networks and the law of diminishing marginal returns: evidence from modeling and human electroencephalographic recordings[J]. 2012.
[35] 王皓月. 基于链接预测的同质开源项目推荐[D].华东师范大学,2022.DOI:10.27149/d.cnki.ghdsu.2022.003970.
[36] Xia X, Zhao S, Zhang X, et al. Understanding the Archived Projects on GitHub[C]//2023 IEEE International Conference on Software Analysis, Evolution and Reengineering (SANER). IEEE, 2023: 13-24.
[37] 赵生宇. 基于 OpenRank 的开源项目内开发者贡献评价[EB/OL].2022. https://blog.frankzhao.cn/openrank_in_project/